\documentclass[english,aps,prb,superscriptaddress,reprint,longbibliograhy]{revtex4-2}
\usepackage[T1]{fontenc}
\usepackage{amsmath}
\usepackage{amssymb}
\usepackage{amsfonts}
\usepackage{bbold}
\usepackage{tikz}
\usepackage{float}
\usepackage{xfp}
\usepackage{tensor}

\usepackage{color}

\usepackage{graphicx}
\usepackage{dcolumn}
\usepackage{bm}
\usepackage{commath}
\usepackage{upgreek}
\usepackage{siunitx}
\usepackage{physics}
\usepackage[english]{babel}
\usepackage[colorlinks=true,linkcolor=cyan,citecolor=blue,urlcolor=cyan]{hyperref}
\usepackage[bottom=2cm,top=2cm, left=1.5cm, right=1.5cm]{geometry}
\usepackage{booktabs}
\usepackage{comment}

\colorlet{linkequation}{blue}
\usepackage[normalem]{ulem}
\usepackage{soul}
\newcommand{\showcomments}{\long\def\comm##1\commend{##1}}

\showcomments

\begin{document}

\title{Magnon chiral damping beyond the Dzyaloshinskii-Moriya interaction}

\author{Darpa Narayan Basu}
\altaffiliation{Present address: Department of Physics and Research Center OPTIMAS,
Rheinland-Pf\"alzische Technische Universit\"at Kaiserslautern-Landau, 67663 Kaiserslautern, Germany}
\email[]{darpa.narayan.basu.physics@gmail.com}
\affiliation{Department of Physics, Indian Institute of Technology (Indian School of Mines), Dhanbad 826004, India}

\author{Subhadip  Ghosh}
\affiliation{Department of Physics, Indian Institute of Technology (Indian School of Mines), Dhanbad 826004, India}

\author{Peter M. Oppeneer}
\affiliation{Department of Physics and Astronomy, Uppsala University, Box 516, SE-75120 Uppsala, Sweden
}

\author{Alexander Mook}
\affiliation{University of M\"unster, Institute of Solid State Theory, 48149 M\"unster, Germany}

\author{Ritwik Mondal}
\altaffiliation{ritwik@iitism.ac.in}
\affiliation{Department of Physics, Indian Institute of Technology (Indian School of Mines), Dhanbad 826004, India}

\date{\today}

\begin{abstract}
Magnon chiral damping denotes damping that is asymmetric with respect to the magnon wave vector. 
Such nonreciprocity has been attributed to antisymmetric Dzyaloshinskii-Moriya spin interactions. 
Here, we provide an alternative theoretical explanation of
magnon chiral damping
using linear spin-wave theory and antisymmetric Gilbert damping introduced 
within the Landau-Lifshitz-Gilbert equation of motion. 
First, employing a minimal model of a single sublattice ferromagnet on a rectangular lattice, we show that the effective magnon damping is independent of the system’s energy, including contributions from Dzyaloshinskii-Moriya interactions. Considering scalar nonlocal Gilbert damping, the effective damping parameter $\alpha_{\rm eff}$ becomes wave-vector dependent. In the long-wavelength limit, $\alpha_{\rm eff}$ acquires a quadratic dependence on $\bm{k}$. 
Furthermore, the antisymmetric component of the Gilbert damping tensor gives rise to asymmetric (nonreciprocal) magnon damping. We find the asymmetry to be linear in $\bm{k}$ in the long-wavelength limit, consistent with recent experiments.
Extending the analysis to a two-sublattice system, we consider cross-sublattice Gilbert damping. In the presence of only scalar Gilbert damping, broken reciprocity in magnon relaxation arises from Dzyaloshinskii-Moriya interactions. In contrast, the cross-sublattice antisymmetric component of the Gilbert damping tensor can generate nonreciprocal magnon damping 
even if the Dzyaloshinskii-Moriya interaction is parametrically set to zero, providing an alternative mechanism for asymmetric spin-wave relaxation. Notably, the nonreciprocity induced by antisymmetric Gilbert damping is roughly an order of magnitude larger than that arising from Dzyaloshinskii-Moriya interactions.

\end{abstract}

\maketitle

\section{Introduction}
 {Magnons are the quantized collective excitations of electron spins in an ordered magnetic material and play a pivotal role in spintronic and magnonic devices \cite{Magnon_quantum_information, chumak2014magnon}. Classically, they can be thought of as a spin wave, with each individual spin carrying out a precession around the effective magnetic field caused by its neighbors.} The field of magnonics focuses on the generation, manipulation, and detection of spin waves for information processing. Unlike conventional electronics, which rely on the transport of electric charge, magnonic systems utilize the intrinsic spin angular momentum of magnons to encode, transmit, and process information. Because magnons can propagate without charge motion, magnonic devices can significantly reduce Joule heating and energy dissipation. As a result, magnonics has emerged as a promising platform for next-generation computing technologies, offering advantages such as low power consumption, high operational speeds, and the potential for high integration densities \cite{Spintronics_review_1, Spintronics_review_2, Chumak2015, Barman_2021, Petti_2022, LENK2011107, Pirro2021}. 

One of the fundamental properties of magnons is that they decay.
For example, the damping of the magnon at the Brillouin-zone center is characterized by the linewidth seen in ferromagnetic resonance measurements \cite{Michael_Farle_1998,Di2015,Brandao2017,Chaurasiya2016,Yu2020MagnonTrap,Yu2020PRB}. Recent studies have demonstrated that magnon damping can exhibit pronounced nonreciprocity, resulting in asymmetric attenuation for counter-propagating spin waves, particularly in systems with interfacial Dzyaloshinskii-Moriya interaction (DMI) \cite{Kai2015PRL,Moon2013,Cortes-Ortuno2013,Zakeri2010,Nembach2015,Yu2013,Cho2015,Safeer2022NatCommun}. The interfacial DMI arises at interfaces between a ferromagnet and a heavy metal (such as Pt, W, or Ta) due to spin-orbit coupling \cite{Yang2015PRL,Fert1980,Kuepferling2023RMP,Belmeguenai2015,Tachhi2017}. This spin interaction not only modifies the spin-wave dispersion relation but also can introduce chirality to the spin-wave modes, leading to direction-dependent dynamics. As a result, magnons traveling in one direction may face enhanced damping, while those moving in the opposite direction experience reduced damping \cite{LIN2026,Xu2026NonreciprocalAFM}.  
Remarkably,
recent studies have demonstrated that even in systems with negligibly small interfacial DMI, such as CoFeB, magnons can still acquire this nonreciprocity \cite{Kim2024PRL},
which has been atributed to $k$-dependent Gilbert damping instead of DMI.
However, the influence of antisymmetric or chiral Gilbert damping has not yet been established.

Both interfacial DMI and chiral Gilbert damping can originate from the structural inversion asymmetry (SIA) of a magnetic system, and they share the dependence on the spin-orbit coupling. Yet, being two different contributions, microscopics can parametrically favor one of them. For example, increasing the thickness of a sample, the contribution of DMI decreases, whereas the magnon chiral damping can become significantly enhanced \cite{Kim2024PRL}. Hence, a sizable magnitude of one does not necessarily imply a significant magnitude of the other~\cite{Nembach2015,Caretta2020,Kuepferling2023RMP,Akanda2020,Kumar2020, Akosa_2024}. Indeed, several experimental and theoretical studies have identified chiral damping as an interface-induced effect that can exist independently of a detectable interfacial DMI contribution to the magnon spectrum \cite{Berrai2026,Rana2020,Akosa2016,Jue2016,Sarkar2026,Ye2024,Pal2025,Kim2016,Freimuth2017}. However, the appearance of magnon chiral damping in the absence of DMI remains largely unexplored. Moreover, 
an effective theoretical framework of the respective contributions of DMI and chiral Gilbert damping to nonreciprocal magnon damping is still lacking. 

In this work, we develop a theoretical framework for investigating magnon chiral damping starting from the Landau-Lifshitz-Gilbert (LLG) equation, with the DMI entering as a parameter.
We extend the LLG dynamics by considering
a Gilbert damping tensor, where the chiral damping naturally emerges
from its antisymmetric component. Nonmagnetic atoms are assumed to induce SIA, thereby allowing both the DMI and chiral Gilbert damping. Further, we consider scalar nonlocal damping contributions \cite{Bhattacharjee2012,Thonig2018,Nagyfalusi2024,Weissenhofer2025}. Within the linear spin-wave theory, we compute the corresponding magnon dispersion relations. 
As a concrete example, we consider a single-sublattice ferromagnet on a rectangular lattice and calculate the magnon dispersion and effective damping ($\alpha_{\rm eff}$) analytically. 
The evaluated expression of $\alpha_{\rm eff}$ shows that it is independent of the DMI for a single-sublattice ferromagnet. Our results further indicate that effective damping becomes $\bm{k}$-dependent when scalar nonlocal damping $\alpha_{ij}$ (for $i \neq j$) is included. Within the long-wavelength approximation, we further derive the analytical expression for effective damping as $\alpha_{\rm eff} = A - Bk^2$, where $A$ and $B$ are constants, reflecting the influence of nonlocal dissipation mechanisms. Such $\bm{k}$-dependent effective magnon damping has been observed experimentally in metallic ferromagnets using spin-wave resonance measurements \cite{k-dep-gilbert}. Note that $\alpha_{\rm eff}$ remains symmetric across the Brillouin zone. In contrast, chiral Gilbert damping breaks this symmetry, producing direction-dependent effective magnon damping even without the scalar nonlocal damping. We find that in the long-wavelength limit this asymmetry varies linearly with wave vector which is consistent with 
experimental findings~\cite{Kim2024PRL}.

To describe the experimental results presented in Ref.\ \cite{Kim2024PRL} for the CoFeB magnetic system, we explicitly compute the effective damping of two-sublattice ferromagnets on a rectangular lattice. In the long-wavelength limit, our proposed model becomes valid to understand  
the experiment.
Introducing cross-sublattice scalar and chiral Gilbert damping terms as nonlocal contributions, and restricting to the nearest neighbors only, we compute the effective damping numerically. 
Our results clearly demonstrate that the effective damping becomes asymmetric within the Brillouin zone when cross-sublattice antisymmetric Gilbert damping is included, even if the DMI is parametrically set to zero.
We further compute the asymmetry parameter, defined as the difference between the effective damping for opposite wave vectors, and find that it vanishes in the absence of both antisymmetric Gilbert damping and DMI.
In the long-wavelength limit, the asymmetry parameter varies linearly with $\bm{k}$, a signature that has been experimentally observed in a heterostructured system where there is negligible DMI in the system \cite{Kim2024PRL}. These results confirm that antisymmetric Gilbert damping plays a significant role as the origin of the observed asymmetry. In the two-sublattice ferromagnet, DMI also breaks the symmetry of the effective damping; however, we find that antisymmetric Gilbert damping induces a much stronger directional asymmetry ($\sim$25 times) compared to DMI within the considered parameter range.    

This paper is organized as follows. In Sec.~\ref{sec1}, we introduce the Landau-Lifshitz-Gilbert spin dynamical equation with isotropic, anisotropic, and antisymmetric Gilbert damping. In Sec.~\ref{Sec2}, a minimal model is presented to obtain the magnon dispersion relation and effective damping for a one-sublattice ferromagnet using linear spin-wave theory. In Sec.~\ref{Sec3}, we extend the analysis to a two-sublattice rectangular lattice to verify the experimentally observed chiral damping in CoFeB. Finally, the main conclusions are presented in Sec.~\ref{Sec4}.      

\section{Landau-Lifshitz-Gilbert equation}\label{sec1}
The Landau-Lifshitz-Gilbert (LLG) equation for a unit spin vector $\bm{\sigma}$ incorporating the Gilbert damping 
 tensor is expressed as \cite{Thonig2014, Mondal2016, Bajpai2019,ZhangPRL2009,Brinker_2022}
 \begin{align}
    \frac{\partial \bm{\sigma}_i}{\partial t} = -\gamma_i{\bm{\sigma}_i}\times\bm{B}^{\rm eff}_i + \bm{\sigma}_i \times \bigg[ \mathcal{D}_{ij} \cdot \frac{\partial \bm{\sigma}_j}{\partial t}\bigg]\,,
    \label{Eq1}
\end{align}
where {\it i} and {\it j} denote the lattice indices. The first term explains the precessional torque that describes how the magnetization vector $\bm{\sigma}$ precesses around the effective magnetic field $\bm{B}^{\rm eff}_i = -\frac{\partial \mathcal{H}}{\partial \bm{\sigma}_i}$, where $\mathcal{H}$ represents the total energy of the system under consideration. The angular frequency of the precession is given by $\gamma\vert\bm{B}^{\rm eff}\vert$. The second term accounts for the loss of energy over time, causing the magnetization to relax and eventually align with the effective magnetic field. Note that the traditional Gilbert damping has often been used as a scalar and isotropic parameter  $\alpha$ \cite{Beaurepaire1996,ostler12,nowak2007handbook,Skubic_2008,Schoen2016,ostler12,Blank2021THz,H_eff_evans}.  
Here, we consider the full damping tensor $\mathcal{D}_{ij}$ \cite{Dhali_2024,Nagyfalusi2025}. Such a tensor, as introduced in Eq.~(\ref{Eq1}), has been derived using several microscopic approaches, including the relativistic Dirac--Kohn--Sham formalism~\cite{Mondal2016,Mondal2017Nutation,Mondal2018} and the breathing Fermi surface theory~\cite{kambersky76}. More recently, additional microscopic theories have further elucidated the origin of tensorial Gilbert damping~\cite{Osorio2024,Pervishko_2022,Thonig2014}. This tensor accounts
for local ($i = j$) and nonlocal Gilbert damping ($i \ne j$).

In general, the damping tensor can be decomposed into isotropic symmetric, 
chiral antisymmetric, and Ising-like anisotropic contributions. In our case, we obtain the scalar nonlocal, antisymmetric, and anisotropic Gilbert damping contributions via $\mathcal{D}_{ij}^{\alpha\beta} = \alpha_{ij}^{\alpha\beta}\delta_{\alpha\beta} + \varepsilon^{\alpha\beta\gamma}\mathcal{C}_{ij}^{\gamma} + \mathcal{I}_{ij}^{\alpha\beta}$, where $\alpha, \beta, \gamma \in \{x,y,z\}$.  
 Incorporating such tensorial damping, the 
LLG equation can be recast in a generalized form as \cite{Mondal2016, Dhali_2024} 
\begin{align}    \frac{\partial \bm{\sigma}_i}{\partial t} &= -\gamma{\bm{\sigma}_i}\times\bm{B}^{\rm eff}_i + \alpha_{ij} \bm{\sigma}_i  \times  \frac{\partial \bm{\sigma}_j}{\partial t}\nonumber
    \\&-\bm{\sigma}_i \times \bigg[\left( \bm{\mathcal{C}}_{ij} \times \frac{\partial \bm{\sigma}_j}{\partial t}\right) - \left({\mathcal{I}}_{ij} \cdot \frac{\partial \bm{\sigma}_j}{\partial t}\right)\bigg]\,. 
    \label{Eq2}
\end{align}
Here, $\alpha_{ij}$ with  $j = i$ is the traditional scalar Gilbert damping parameter $\alpha _{ii} = \alpha_i$. 
Further, $\alpha_{ij}$ with  $j \neq i$ denotes the scalar nonlocal Gilbert damping  \cite{Bhattacharjee2012,Thonig2018,Osorio2024}. Nonlocal damping arises from the coupling of spin dynamics at different lattice sites via conduction electrons \cite{Weissenhofer2025} or spin currents, where spin pumping, electron-mediated interactions, and inter-sublattice coupling enable the transfer of angular momentum across the system~\cite{Lu2023PRB}.

The vector $\bm{\mathcal{C}}_{ij}$  denotes the chiral or antisymmetric damping contribution. Within the microscopic formalism of Gilbert damping derived by Bhattacharjee {\it et al.}~\cite{Bhattacharjee2012}, the nonlocal damping tensor can acquire antisymmetric contributions in systems with broken inversion symmetry and spin-orbit coupling. These damping terms obey the same symmetry constraints as the   DMI and may therefore be regarded as its dissipative analogue~\cite{Kim2013,D-Moriya-original,Gueneau2026}.  
On the other hand,
${\mathcal{I}}_{ij}$ is the anisotropic Ising-like damping tensor contributions. Throughout this work, we restrict our analysis to the first three terms in Eq.~(\ref{Eq2}), neglecting ${\mathcal{I}}_{ij}$. 
This approximation is justified for two reasons. First, the diagonal components $\mathcal{I}_{ii}$
contribute in a manner similar to the isotropic Gilbert damping term $\alpha_i$. Second, the off-diagonal components are typically much smaller than the diagonal ones and therefore contribute negligibly~\cite{Dhali_2024}. 

In the following, we perform linear spin-wave theory calculations for both single- and two-sublattice ferromagnets. In particular, for two-sublattice systems, we incorporate inter-sublattice scalar and antisymmetric Gilbert damping, which accounts for the nonlocal contribution~\cite{Kamra2018}.

\section{Single Sublattice Ferromagnet}
\label{Sec2}

\subsection{Linear Spin-Wave Theory}
We consider a two-dimensional magnetic system in the $xz$-plane on a rectangular lattice. The magnetic properties are described by the classical spin Hamiltonian
\begin{align}
\mathcal{H}
&= \frac{1}{2}\sum_{i\neq j}\bm{\sigma}_i \bm{J}_{ij} \bm{\sigma}_j
+ \sum_i K^x(\sigma_i^x)^2
- \sum_i M_i\,\bm{B}_i\cdot\bm{\sigma}_i,
\label{Eq4}
\end{align}
where $\bm{\sigma}_i$ is a unit vector representing the direction of the magnetic moment at lattice site $i$, $\bm{J}_{ij}$ represents the exchange interaction tensor~\cite{H_eff_evans}, $K^x$ denotes the uniaxial anisotropy constant with the easy axis along the ${\hat{\bm x}}$ direction, $\bm{B}_i$ is the external magnetic field, and $M_i$ is the magnetic moment of atom $i$. The diagonal elements of the exchange tensor, namely $J_{ij}^{xx}$, $J_{ij}^{yy}$, and $J_{ij}^{zz}$, describe the Heisenberg exchange interactions, whereas the off-diagonal elements give rise to the  
DMI, with
$D_{ij}^{\alpha}=\varepsilon^{\alpha\beta\gamma}J_{ij}^{\beta\gamma}$,
where $\alpha,\beta,\gamma\in\{x,y,z\}$ and $\varepsilon^{\alpha\beta\gamma}$ is the Levi--Civita tensor~\cite{Rozsa_2013,D-Moriya-original}.  
Equation~(\ref{Eq4}) describes the magnetic system for arbitrary spin configurations. 

Throughout this work, we assume that {the classical ground state is a fully polarized ferromagnetic configuration with all magnetic moments aligned along the ${x}$ direction, corresponding to the easy axis of the system. This is possible if the uniaxial anisotropy is sufficiently strong and negative ($K^x<0$) or the magnetic field along the ${x}$ direction is sufficiently high.}
To investigate small-amplitude spin-wave excitations, we expand the spin vectors about this equilibrium configuration. The spin direction at site $i$ is parametrized in terms of two small transverse deviation variables, $\beta_{1i}$ and $\beta_{2i}$, as~\cite{Rozsa_2013,Mondal2022,Ghosh2026}
\begin{align}
\bm{\sigma}_i =
\begin{pmatrix}
1-\dfrac{\beta_{1i}^{2}}{2}-\dfrac{\beta_{2i}^{2}}{2}\\
\beta_{2i}\\
-\beta_{1i}
\end{pmatrix},
\label{Eq3}
\end{align}
where terms beyond second order in the deviation variables have been neglected. Substituting Eq.~(\ref{Eq3}) into the Hamiltonian (\ref{Eq4}) and retaining terms up to quadratic order in $\beta_{1i}$ and $\beta_{2i}$ yields the harmonic spin-wave Hamiltonian, which forms the basis of the subsequent analysis.

We compute the effective fields via $\bm{B}^{\rm eff}_i = -\frac{1}{M_i}\frac{\partial \mathcal{H}}{\partial \bm{\sigma}_i}$ and insert these into the LLG dynamics described by Eq.\ (\ref{Eq2}). Further, we linearize the LLG equation 
and introduce spatial Fourier transformations of  the angle variables $\tilde\beta_{1}(\bm{k})$ and $\tilde\beta_{2}(\bm{k})$. Using a circular basis $\tilde\beta_{\pm}(\bm{k}) = \tilde\beta_2(\bm{k}) \pm {\rm i}\tilde\beta_1(\bm{k})$, the LLG dynamics in Eq.\ (\ref{Eq2}) can be recast as
\begin{align}
\left[\pm {\rm i}\omega {\mathcal{A}({\bm k})} + {\mathcal{J}}({\bm k})\right]\tilde\beta_{\pm}(\bm{k}) = B_{\pm}(\bm{k})\,.
\label{Eq5}
\end{align}
Here, we used $B_{\pm}({\bm k}) = B_z({\bm k}) \pm {\rm i} B_y({\bm k})$, and assumed $\tilde{\beta}_{\pm} \propto e^{\pm {\rm i}\omega t}$.   
${\mathcal{A}({\bm k})}$ and ${\mathcal{J}({\bm k})}$ capture information on damping and magnetic interactions present in the system, respectively, and are given by the following expressions
\begin{align}
    \mathcal{A}(\bm{k}) &= 1 \pm {\rm i}\alpha \pm {\rm i}\tilde{\alpha}(\bm{k}) - \tilde{C}(\bm{k}), \\
    \mathcal{J}(\bm{k}) &= \frac{\gamma}{M} \bigg[\pm {\rm i}\tilde{J}^{zz}(\bm{k}) \mp {\rm i} \tilde{\Omega} -\tilde{D}(\bm{k}) \bigg].
\end{align}
The detailed derivations with their expressions are provided in Appendix \ref{AppendixA}.
\begin{figure}[tbh!]
    \centering
    \includegraphics[width=1\linewidth]{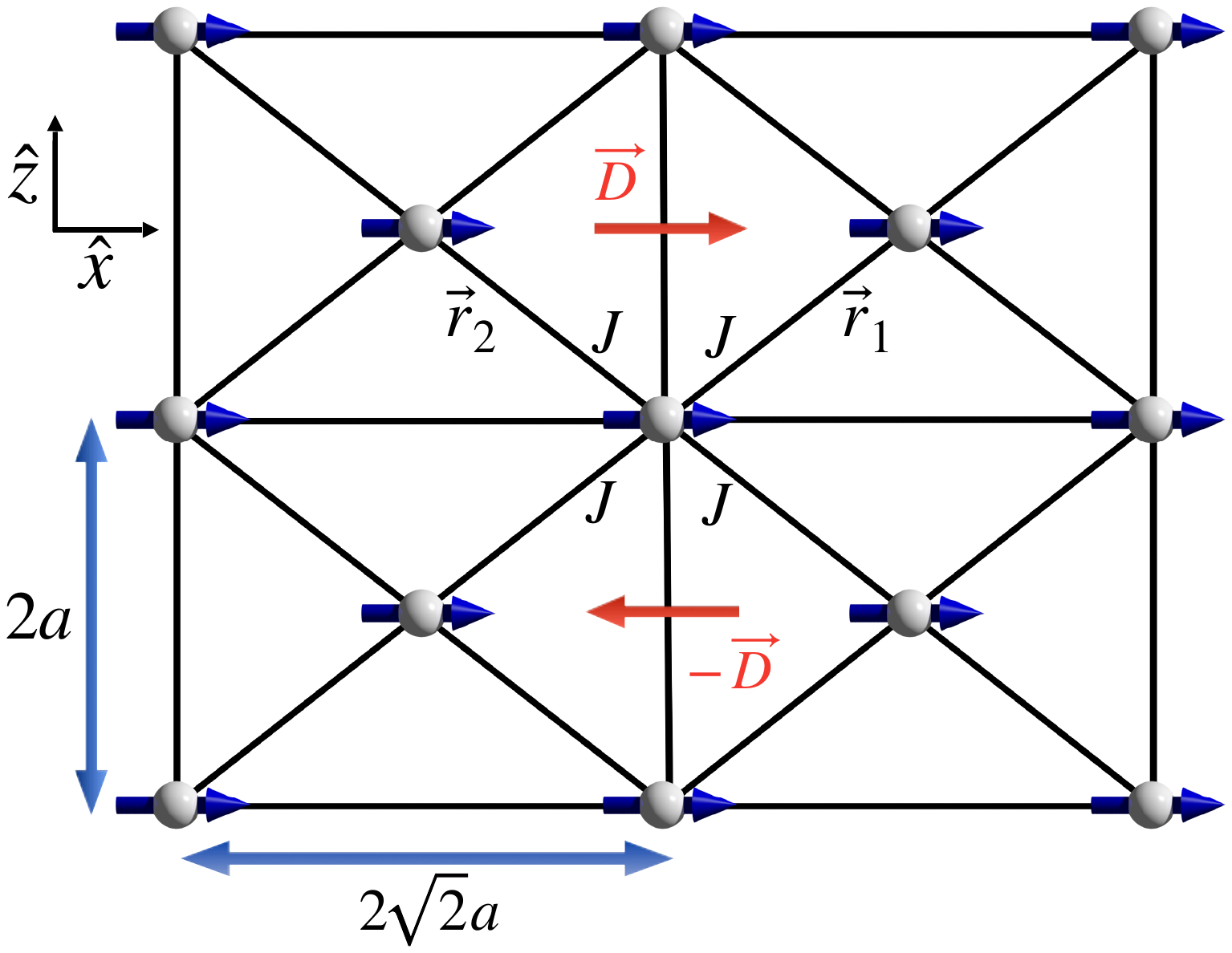}
    \caption{A sketch of the ground state of a two-dimensional single-sublattice ferromagnet. All spin vectors are represented by blue arrows. $\bm{r}_1$ and $\bm{r}_2$ denote the primitive cell lattice vectors. The nearest-neighbor exchange interaction is denoted by $J$. The red arrows indicate the direction of the DM vector ($\bm{D}$) for bonds connecting the central atom to its second-nearest neighbors. Nonmagnetic atoms are assumed to induce SIA, thereby allowing an in-plane DMI and chiral Gilbert damping.}
   \label{fig:Fig1}
\end{figure}

\subsection{Effective Damping}
We consider a single-sublattice ferromagnet, as sketched in Fig.~\ref{fig:Fig1}. The lattice structure resembles an Fe/W(110) monolayer \cite{Udvardi2009}. The lattice lies in the $xz$ plane, with the ${\bm{\hat x}}$ axis defining the equilibrium magnetization direction. We choose the lattice vectors as ${\bm r}_1 = \sqrt{2}a\hat{\bm x} + a\hat{\bm z}$ and ${\bm r}_2 = -\sqrt{2}a\hat{\bm x} + a\hat{\bm z}$, with $2a$ and $2\sqrt{2}a$ as lattice constant along ${\bm{\hat z}}$ and ${\bm{\hat x}}$ directions, respectively. The magnetic exchange interactions are assumed to act only between first-nearest neighbors. Owing to the SIA, the DMI vectors are taken for the central atom to its second-nearest neighbors, as illustrated in Fig.~\ref{fig:Fig1} \cite{D-Moriya-original}. Considering all these, the Fourier-transformed exchange and DMI terms are evaluated.

In our model, the local Gilbert damping is taken to be uniform, 
 $\alpha_i = \alpha$, while the scalar nonlocal damping is assumed to be $\alpha_{ij} = \alpha^\prime$, restricted to first nearest neighbors. This leads to a 
$\bm{k}$-dependent effective Gilbert damping. The antisymmetric Gilbert damping is taken in accordance with the symmetry of the 
DMI vectors as both share the same symmetry constraints~\cite{Akosa2016, Nagaosa2013, Thiaville_2012}. Thus, the antisymmetric Gilbert damping vectors are $\bm{\mathcal{C}}_{ij} = \pm C^\prime\hat{\bm{x}}$ which also acts between the second nearest neighbors only~\cite{Kim2024PRL,D-Moriya-original}. Note that although both the antisymmetric Gilbert damping and DMI share the same symmetry constraints, both are governed by different physical mechanisms~\cite{Kim2024PRL,Akosa_2024,Gunggord2016}. Thus, we can treat both of them as independent parameters in our model. 
The expressions obtained for scalar nonlocal damping and antisymmetric damping are 
\begin{align}
\tilde{\alpha}(\bm{k}) &= 4 \alpha'\Big[\cos(\sqrt{2}k_xa) \cos(k_za)\Big]\,,\\
    \tilde{C}(\bm{k}) &= 2{\rm i}C'\sin(2k_za)\,.
\end{align}
Using these expressions and setting $B_y = B_z = 0$, we solve Eq.~(\ref{Eq5}) numerically to obtain the complex magnon frequency 
$\omega(\bm{k}) = {\tt Re}\left[\omega(\bm{k})\right] + {\rm i}{\tt Im}\left[\omega(\bm{k})\right]$. The real part determines the dispersion, whereas the imaginary part encodes the damping. The effective magnon damping is computed via  
$\alpha_{\rm eff}(\bm{k}) = | {{\tt Im}\left[\omega(\bm{k})\right]/{\tt Re}\left[\omega(\bm{k})\right]} |$ as,
\begin{equation}
    \alpha_{\rm eff}(\bm{k}) = \alpha + \tilde{\alpha}(\bm{k}) \mp {\tt Im} \big[\tilde{C}(\bm{k})\big]\,.
    \label{Eq11}
\end{equation}
This is one of the central results of the article. This expression demonstrates that the effective damping acquires a momentum dependence arising from scalar nonlocal and from antisymmetric contributions. In particular, the term involving ${\tt Im}\big[\tilde{C}(\bm{k})\big]$ introduces an asymmetry in $\alpha_{\rm eff}(\bm{k})$, leading to nonreciprocal magnon damping for opposite momenta.
Restricting to scalar nonlocal Gilbert damping $\alpha^\prime$ and ignoring the antisymmetric Gilbert damping contributions, Eq.~\eqref{Eq11} simplifies in the long-wavelength limit to the form
\begin{equation}
    \lim _{\bm{k}\rightarrow 0}\alpha_{\rm eff}(\bm{k}) \Big \vert_{C^\prime = 0}\approx (\alpha+ 4\alpha^\prime) - 4\alpha^\prime a^2 k_x^2- 2\alpha^\prime a^2 k_z^2 \,.
    \label{Eq12}
\end{equation}
This result shows that the effective damping acquires a quadratic dependence on the wave vector, reflecting the nonlocal nature of the dissipative processes. Such behavior is consistent with the experimental findings reported by Li \textit{et al.}, confirming the validity of the nonlocal damping description  
for which the damping correction scales as $\Delta \alpha \sim A_k k^2$~\cite{k-dep-gilbert}.  
More generally, our expression captures possible anisotropic contributions arising from the underlying rectangular lattice symmetry. This effective damping remains notably symmetric in $\bm{k}$ without the antisymmetric contributions, as it is evident that $\tilde{\alpha}(\bm{k}) = \tilde{\alpha}(-\bm{k})$, even in the short-wavelength limit.  
\begin{figure*}
    \centering
\includegraphics[width=0.7\linewidth]{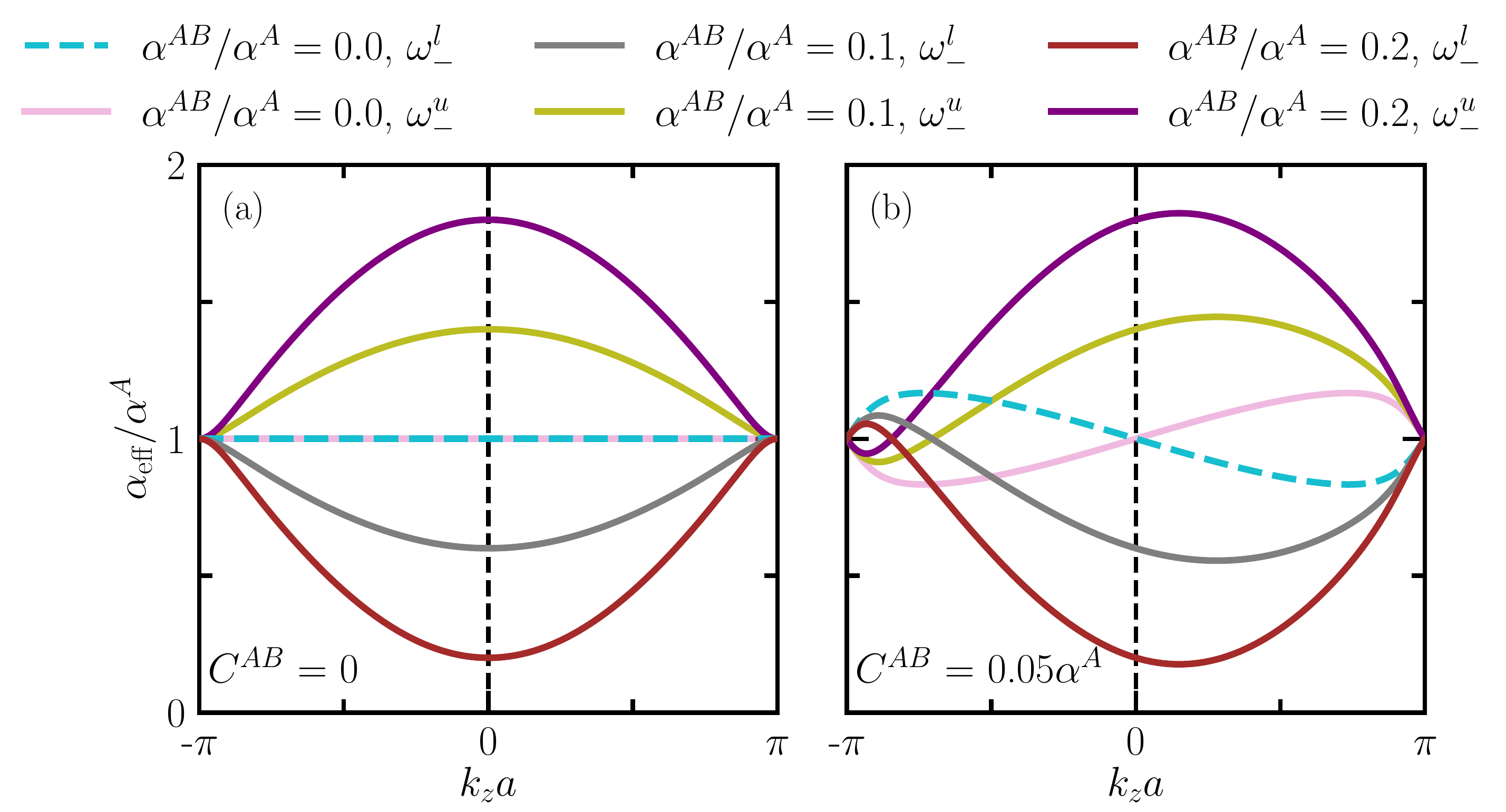}
\caption{Plot of the effective magnon damping ($\alpha_{\rm eff}$) versus scaled wave vector $k_za$, setting $k_x = 0$, for (a) without, and (b) with the antisymmetric Gilbert damping ($C^\prime$) included. In both panels, the dependence on the nonlocal Gilbert damping parameter ($\alpha^\prime$) is shown through the line colors.
}
\label{fig:Fig2}
\end{figure*}

Now we consider the magnons propagating perpendicular to the anisotropy axis $\hat{\bm{z}}$, which is similar to the Damon-Eshbach mode~\cite{Damon1960,Sun2025,Wang2016}. Incorporating the antisymmetric Gilbert damping term in Eq.~\eqref{Eq11} as well, in the long-wavelength limit $\alpha_{\rm eff} (\bm{k})$ takes the following form
\begin{equation}
    \lim _{\bm{k}\rightarrow 0}\alpha_{\rm eff} (\bm{k}) \approx (\alpha+ 4\alpha^\prime) - 2\alpha^\prime a^2 k_z^2 \mp 4C^\prime a k_z \, .
    \label{Eq14n}
\end{equation}
This equation demonstrates that in the presence of antisymmetric damping only,  $\alpha_{\rm eff}$ acquires a linear dependence on $\bm{k}$ in the long-wavelength limit. To quantify the asymmetry in the effective damping, we define the following asymmetry parameter
\begin{align}
\Gamma(\bm{k}) = \frac{\alpha_{\rm eff}(\bm{k}) - \alpha_{\rm eff}(-\bm{k})}{\alpha}\,.
\label{Eq15n}
\end{align}
We find that in the long-wavelength limit, $\Gamma(\bm{k}) \propto k_z$ which is consistent with the experimental observations~\cite{Kim2024PRL}.

Next, we plot the effective damping according to Eq.~(\ref{Eq11}) in Fig.~\ref{fig:Fig2} by parametrically setting antisymmetric Gilbert damping to zero or nonzero.
We present the variation of $\alpha_{\rm eff}(\bm{k})$ as a function of $k_z a$ while fixing $k_x = 0$. The dependence on $k_x$ is not shown, since, as is evident from the analytical form of $\tilde{C}(\bm{k})$, the antisymmetric damping contribution vanishes along this direction. A careful inspection of Fig.~\ref{fig:Fig2}(a) shows that the inclusion of interatomic Gilbert damping $(\alpha^\prime)$ renders $\alpha_{\rm eff}$ wave-vector dependent. Specifically, it is enhanced at ${\bm k} = 0$ and progressively suppressed toward the Brillouin-zone boundaries
at $k_z = \pm\pi/a$.  At the high-symmetry points $k_z = \pm\pi/2a$, the effective damping remains unchanged, reflecting the vanishing contribution of the nonlocal damping term at these $\bm{k}$ points. Upon introducing the antisymmetric damping $(C^\prime)$, $\alpha_{\rm eff}$ becomes asymmetric in $k_z$, as shown in Fig.~\ref{fig:Fig2}(b), indicating nonreciprocal damping for counter-propagating spin waves. This asymmetry arises from the odd-in-momentum nature of the antisymmetric term, i.e., $\tilde{C}(\bm{k}) = -\tilde{C}(-\bm{k})$, which leads to different damping rates for magnons with opposite wave vectors. 

Although the present model includes DMI in the magnetic energy in Eq.\ (\ref{Eq4}),  the effective damping is primarily governed by dissipative terms and remains, to leading order, independent of the specific form of the magnetic energy contributions. Therefore, the damping behavior obtained here is expected to hold irrespective of the parametrized strength of the DMI. Although both the DMI and antisymmetric Gilbert damping are symmetry-allowed interactions in the system, the DMI can be parametrically set to zero while retaining a finite antisymmetric damping. Consequently, the nonreciprocal magnon damping remains finite even in the absence of a DMI contribution to the Hamiltonian in Eq.~(\ref{Eq4}).

To summarize, our model for a single-sublattice ferromagnet demonstrates that the scalar nonlocal Gilbert damping renders the effective damping parameter $\alpha_{\rm eff}$ wave-vector dependent. In the long-wavelength limit, this contribution is quadratic in $\bm{k}$ and remains symmetric under $\bm{k}\rightarrow-\bm{k}$. In contrast, the nonlocal antisymmetric damping introduces a term that is linear in $\bm{k}$, thereby breaking the $\bm{k}\rightarrow-\bm{k}$ symmetry and giving rise to nonreciprocal magnon damping. The proposed model therefore provides a theoretical framework for understanding recent experimental observations of wave-vector-dependent and asymmetric magnon damping~\cite{k-dep-gilbert,Kim2024PRL}.

\section{Two-sublattice ferromagnet}
\label{Sec3}

\subsection{Linear Spin-Wave Theory}
To investigate the experimental observation in CoFeB \cite{Kim2024PRL}, we further compute the effective damping in a two-sublattice ferromagnet assuming a rectangular lattice.
Although CoFeB is amorphous and lacks long-range crystal structure, our model can provide a possible theoretical framework for the experiment in the long-wavelength limit. This is because in this limit the results are independent of the crystal structure. Moreover, the wavelength of the magnons of the corresponding experiment is much larger than the lattice spacing ($\sim 10^3$ times).

In this model, the unit spin vector at the $i$-th lattice site is denoted by $\bm{\sigma}_{i_p}^p$, where $p \in \{A, B\}$ labels the two sublattices. Owing to the presence of multiple sublattices, the damping parameters $\alpha_{ij}$ and the components of the antisymmetric damping $\bm{\mathcal{C}}_{ij}$ in the LLG equation [Eq.~\eqref{Eq2}] must be treated as $2 \times 2$ matrices in sublattice space \cite{Kamra2018, Mondal2021PRB, Cheng2014, Tserkovnyak2005Nonlocal, Liu2017PRM, Li2022AFM, Troncoso2021, MondalJPCM2021}. Accordingly, the LLG equation for $\bm{\sigma}_{i_p}^p$ can be written as
\begin{align}
    \frac{\partial \bm{\sigma}_{i_p}^p}{\partial t} = &-\gamma_{i_p}^p \bm{\sigma}_{i_p}^p \times \bm{B}^{p}_{{{\rm eff},i_p}} + 
    \alpha_{i_p j_q}^{pq}\bm{\sigma}_{i_p}^p\times \frac{\partial \bm{\sigma}_{j_q}^q}{\partial t} \nonumber\\& - 
    \bm{\sigma}_{i_p}^p \times \left[\bm{\mathcal{C}}^{pq}_{i_p j_q}\times \frac{\partial \bm{\sigma}_{j_q}^q}{\partial t}\right].
\label{Eq13}
\end{align}
 As in the single-sublattice case, we assume that the ground-state spins are aligned along the ${x}$ direction. Small deviations from equilibrium are described by the angular variables $\beta_{1 i_p}^{p}$ and $\beta_{2 i_p}^{p}$ \cite{Rozsa_2013}. The spin Hamiltonian of the system is given by
\begin{align}
    \mathcal{H}  &= \frac{1}{2}
    \sum_{i_p \neq j_q}\bm{\sigma}_{i_p}^p \bm{J}^{pq}_{i_pj_q} \bm{\sigma}_{j_q}^q + 
    \sum_p\sum_{i_p} 
    K^{x,p} (\sigma_{i_p}^{x,p})^2 \nonumber\\& \quad- \sum_p\sum_{i_p}
    \bm{B}_{i_p}\cdot M^p_{i_p} \bm{\sigma}_{i_p}^p.
    \label{Eq14}
\end{align}
Here, $J_{i_pj_q}^{pq}$ denotes the exchange tensor, $K^{x,p}$ corresponds to axial anisotropy along the ${x}$ direction, $\bm{B}_{i_p}$ represents the external magnetic field, and $M^p_{i_p}$ is the magnetization at lattice site $i_p$.
We introduce the spatial Fourier transformations of angle variables as $\tilde\beta_{1,2}^p(\bm{k})$. Using a circular basis, the LLG dynamics in Eq.\ \eqref{Eq13} can be recast in the limit of small angles as [see Appendix \ref{AppendixC} for the detailed calculations]
\begin{align}
\left[\pm {\rm i}\omega\mathcal{A}_2(\bm{k}) + \mathcal{J}_2(\bm{k})\right]\tilde{\bm{\beta}}_{\pm}(\bm{k}) = \mathcal{B}_{\pm}(\bm{k}),
\label{Eq15}
\end{align}
where
\begin{align}
    \bm{\tilde{\bm{\beta}}_{\pm}} &= \left(\tilde\beta_{\pm}^{A} \quad \tilde\beta_{\pm}^{B} \right)^{\bm{T}},
    \\
    \mathcal{B}_{\pm} &= \left(\gamma^A\tilde B_{\pm} \quad \gamma^B\tilde B_{\pm} \right)^{\bm{T}}.
\end{align}
In contrast to the single-sublattice case, $\mathcal{A}_2(\bm{k})$ and $\mathcal{J}_2(\bm{k})$ are $2 \times 2$ matrices in sublattice space. The diagonal elements describe interactions between spins on the same sublattice, while the off-diagonal elements encode interactions between different sublattices. Their explicit forms are provided in Appendix~\ref{AppendixC}.

In the following, we solve Eq.~\eqref{Eq15} numerically to evaluate the $\bm{k}$-dependence of the effective damping. To ensure a realistic description, we employ experimentally reported parameters listed in Table~\ref{table1}. 

\begin{figure}[tbh!]
    \centering
    \includegraphics[width=1\linewidth]{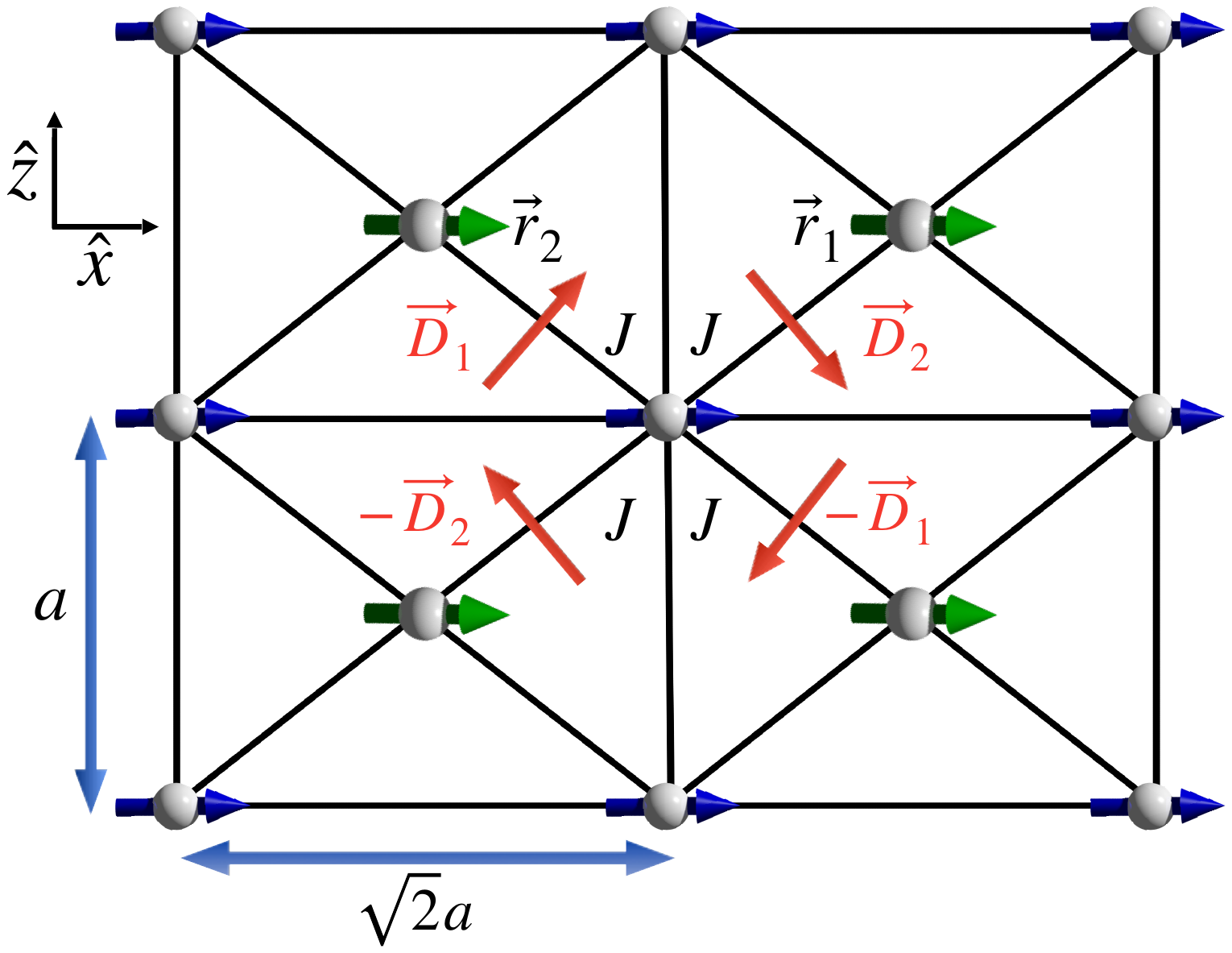}
    \caption{A schematic illustration of the ground state of a two-dimensional ferromagnet. The blue and green spin vectors represent Fe and Co atoms, respectively. The vectors $\bm{r}_1$ and $\bm{r}_2$ denote the sublattice basis vectors. The nearest-neighbor exchange interaction is represented by $J$. The red arrows indicate the directions of the Dzyaloshinskii--Moriya vectors ($\bm{D}_1$ and $\bm{D}_2$).}
    \label{fig:Fig3}
\end{figure}

\subsection{Effective Damping}
\begin{figure*}[tbh!]
    \centering
\includegraphics[width=0.75\linewidth]{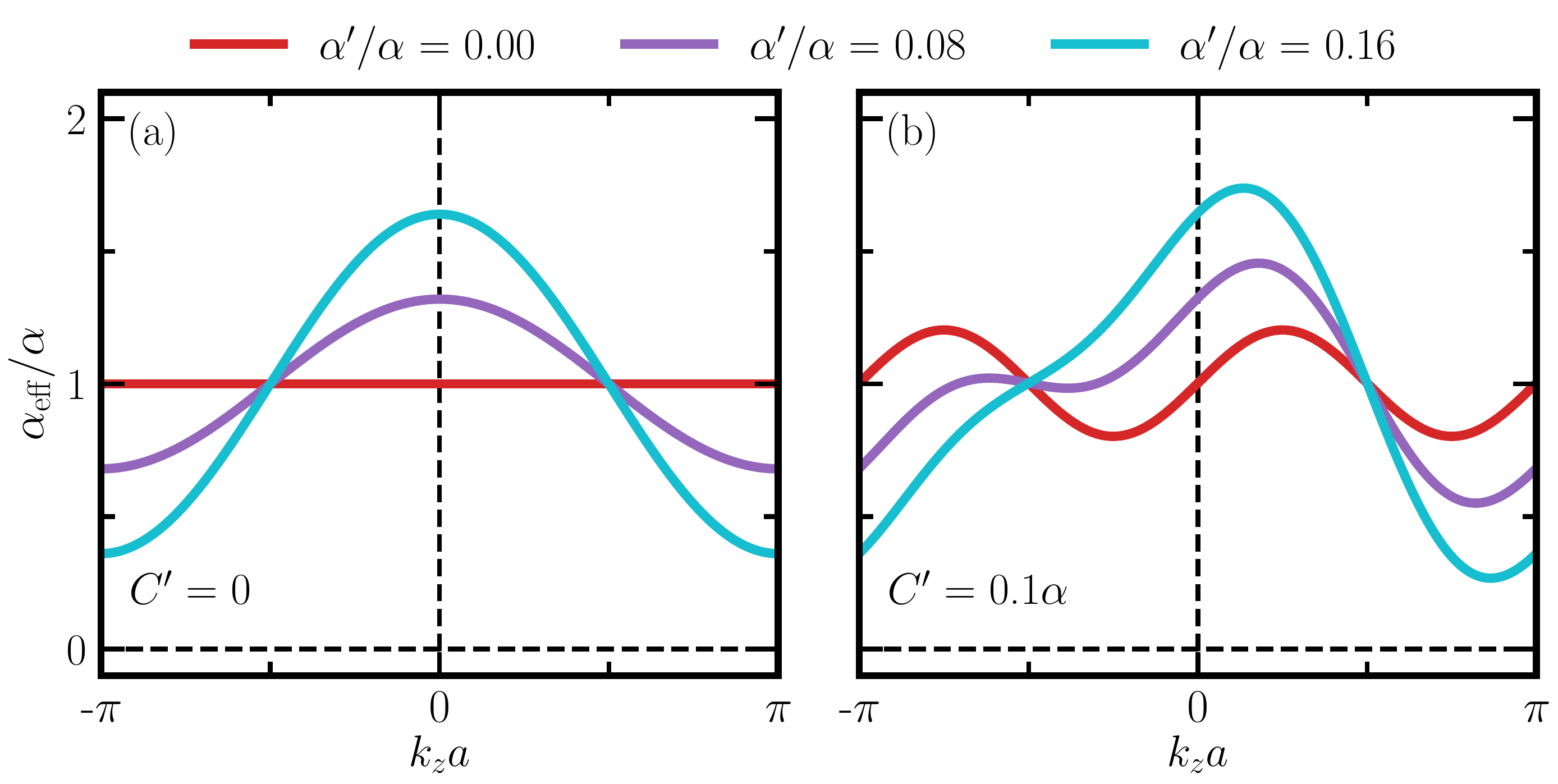}
    \caption{    Calculated effective damping $\alpha_{\rm eff}$  versus $k_za$ (at $k_x=0$), assuming zero DMI, for (a) without antisymmetric damping $C_{AB}$, and (b) with antisymmetric damping. The dependence on the inter-sublattice Gilbert damping $\alpha^{AB}$ is expressed by the line colors.
    Marked variations of $\alpha_{\rm eff}$ with $\alpha^{AB}$ are observed; 
    these vanish at the Brillouin-zone boundaries, $k_z = \pm \pi/a$.}
    \label{fig:Fig4}
\end{figure*}
We consider a two-sublattice ferromagnet, representing CoFeB, as illustrated in Fig.~\ref{fig:Fig3}, where the blue and green spins correspond to Fe and Co atoms, respectively. 
We assume the lattice lies in the $xz$ plane. The primitive lattice vectors are chosen as $\bm{R}_1 = \sqrt{2}a\,\hat{\bm{x}}$ and $\bm{R}_2 = a\,\hat{\bm{z}}$, while the sublattice basis vectors are given by $\bm{r}_1 = \frac{a}{\sqrt{2}}\hat{\bm{x}} + \frac{a}{2}\hat{\bm{z}}$ and $\bm{r}_2 = -\frac{a}{\sqrt{2}}\hat{\bm{x}} + \frac{a}{2}\hat{\bm{z}}$. The Heisenberg exchange interaction is restricted to first nearest neighbors. In contrast to the single-sublattice case, the 
DMI is allowed between first nearest neighbors~\cite{D-Moriya-original,Freimuth2014,Atureri2025}, as shown in Fig.~\ref{fig:Fig3}. The nearest-neighbor bonds connect inequivalent atomic sites (Fe and Co) and therefore lack an inversion center at the bond midpoint. 
Using the lattice symmetries, we determine the orientation of the corresponding DM vectors. 

In the dynamical equation (\ref{Eq15}), we restrict the inter-sublattice scalar Gilbert damping to act within the first nearest neighbors only. The antisymmetric damping $\bm{\mathcal{C}}_{i_pj_q}^{pq}$ vectors are considered to be in the same direction as that of the DM vector. Accordingly, we evaluate the Fourier-transformed inter-sublattice scalar and antisymmetric Gilbert damping terms as,
\begin{align}
\tilde{\alpha}^{AB}(\bm{k}) &= 4\alpha^{AB}\Big[\cos(\frac{k_xa}{\sqrt{2}}) \cos(\frac{k_za}{2})\Big],\\
\tilde{C}^{AB}(\bm{k}) &= 4{\rm i}C^{AB}\Big[\cos(\frac{k_xa}{\sqrt{2}}) \sin(\frac{k_za}{2})\Big].
\end{align}
Here, $\alpha^{AB}$ and $C^{AB}$ denote the cross-sublattice nonlocal and antisymmetric damping strengths, respectively.

\begin{table}
\caption{\label{table1}
Exchange interaction, anisotropy energy, local Gilbert damping, and equilibrium magnetic moments used in the calculations for the two-sublattice ferromagnet (mimicking CoFeB).
}
\begin{ruledtabular}
\begin{tabular}{lcc}
Quantity (unit) & Value & Reference \\[0.05cm] 
\hline 
\\[-0.25cm]
$J^{AB}$ (J) 
& {$-1.547 \times 10^{-20}$} 
& \cite{H-sato-experiment} \\

$K$ (J) 
& {$-1.350 \times 10^{-22}$} 
& \cite{H-sato-experiment} \\

$\alpha^A = \alpha^B$ 
& $8\times 10^{-3}$ 
& \cite{Gilbert-damping-value} \\
$M^A = M^{\rm Fe}$ ($\mu_B$) 
& $1.75$ & \cite{M-hoffherr-exp}\\
$M^B = M^{\rm Co}$ ($\mu_B$) & $2.60$
& \cite{M-hoffherr-exp}\\
$B^x$ (mT) & 300 & \cite{Kim2024PRL}
\end{tabular}
\end{ruledtabular}
\end{table}

We numerically solve Eq.~\eqref{Eq15} and obtain the resonance frequencies $\omega(\bm{k})$. For two-sublattice magnets, we denote these resonance frequencies as $\omega_-^l(\bm{k})$ and $\omega_-^u(\bm{k})$, representing the acoustic and optical branches, respectively. The negative sign in the frequency corresponds to the branch of the solution of Eq.~\eqref{Eq15} obtained with the negative sign, while the positive branch is given by its complex conjugate. Here, we consider the solution of Eq.~\eqref{Eq15} corresponding to the negative sign.  Using the parameters listed in Table \ref{table1}, we further compute the effective damping via $\alpha_{\rm eff}(\bm{k}) = | {{\tt Im}\left[\omega_-(\bm{k})\right]/{\tt Re}\left[\omega_-(\bm{k})\right] | }$ for acoustic and optical branches. 

The computed $\alpha_{\rm eff}$ are shown in Fig.~\ref{fig:Fig4}. Here, we parametrically choose the DMI to be zero. The effective damping is $\bm{k}$-independent when only local Gilbert damping ($\alpha_A = \alpha_B$) is present, and both inter-sublattice and antisymmetric contributions are neglected. As in the single-sublattice case, a finite inter-sublattice scalar damping ($\alpha^{AB} \neq 0$) induces a wave-vector dependence in $\alpha_{\rm eff}$. Away from the Brillouin-zone boundaries at $k_z =\pm \pi/a$, $\alpha_{\rm eff}$ is enhanced (suppressed) for the $\omega_{-}^u$ ($\omega_{-}^l$) mode. Such enhancement (suppression) remains of the same magnitude with respect to the $\bm{k}$-independent mode. 
  
As a result, inter-sublattice damping affects the modes differently, leading to mode-dependent dissipation. Nevertheless, for a given mode, the effective damping remains symmetric under $\bm{k} \rightarrow -\bm{k}$. These findings are consistent with the results obtained for the single-sublattice case discussed earlier. 

Upon including antisymmetric Gilbert damping (i.e., $C^{AB}$), we observe a pronounced asymmetry in the effective damping $\alpha_{\rm eff}$ [Fig.~\ref{fig:Fig4}(b)].
We note that the enhancement (reduction) of the damping away from the zone boundary remains consistent with Fig.~\ref{fig:Fig4}(a) for the $\omega_-^u$ ($\omega_-^l$) mode. However, the inclusion of antisymmetric damping renders the otherwise $\bm{k}$-independent mode $\bm{k}$-dependent and asymmetric across the Brillouin zone. This demonstrates that, even if the DMI is parametrically set to be zero,  antisymmetric damping breaks the $\bm{k} \rightarrow -\bm{k}$ symmetry of magnon decay rates. This provides a theoretical framework for understanding the asymmetric damping reported experimentally~\cite{Kim2024PRL}.
\begin{figure}[tbh!]
    \centering
\includegraphics[width=0.95\linewidth]{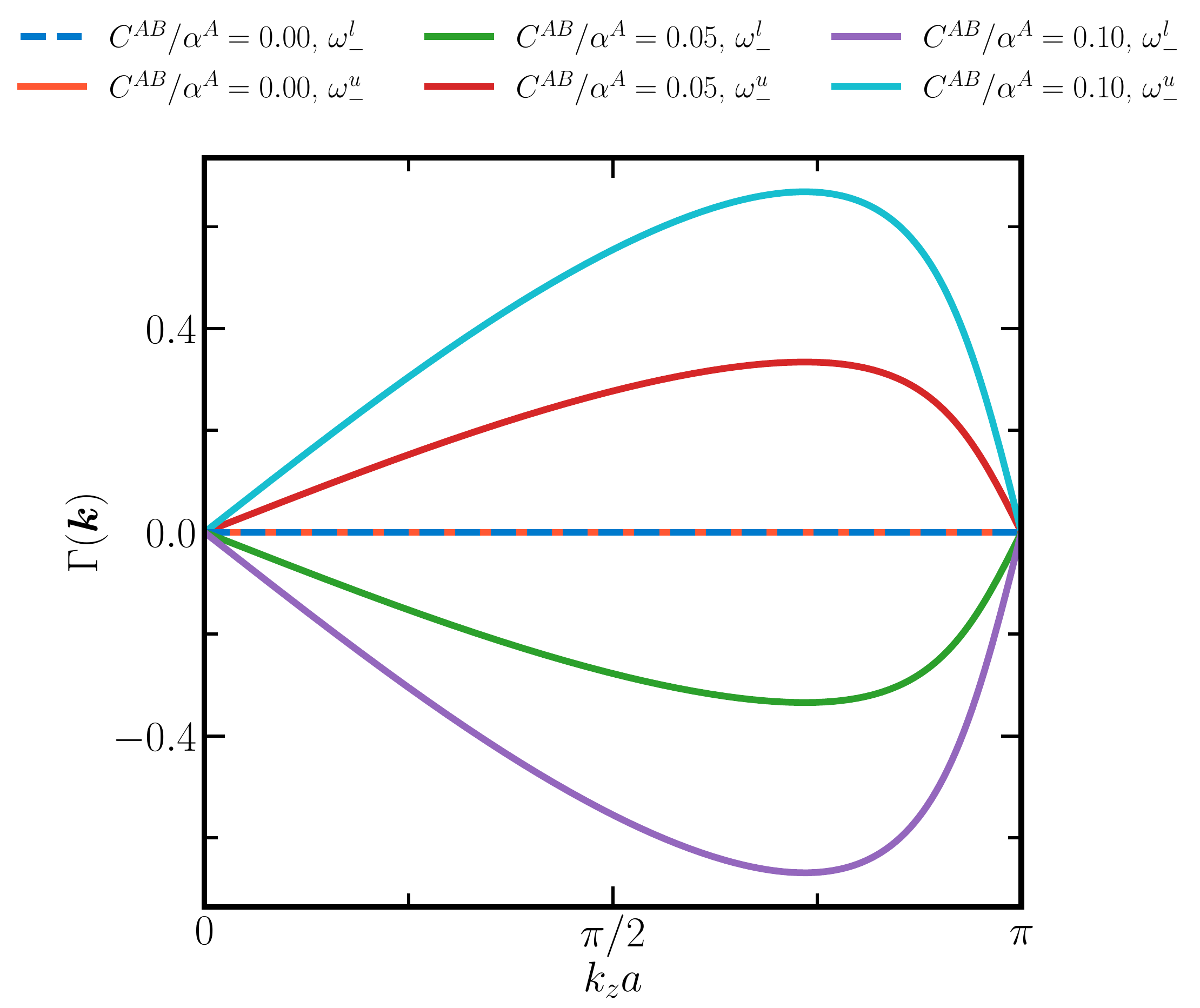}
    \caption{The calculated asymmetry
    $\Gamma(\bm{k})$ in the effective damping $\alpha_{\rm eff}(\bm{k})$  plotted in the first Brillouin zone, setting $k_x = 0$. Line colors  depict the dependence on the cross-sublattice  antisymmetric damping $C^{AB}$. }
    \label{fig:Fig5}
\end{figure}

To quantify the asymmetry of the effective damping, we use Eq.~\eqref{Eq15n} to compute this quantity for $\alpha = \alpha_A = \alpha_B$.
The resulting asymmetry $\Gamma(\bm{k})$ is plotted in Fig.~\ref{fig:Fig5}. We find that $\Gamma(\bm{k})$ vanishes when the cross-sublattice antisymmetric damping is zero ($C^{AB} = 0$), indicating symmetric damping, whereas a finite $C^{AB} \neq 0$ leads to a nonzero asymmetry. This demonstrates that antisymmetric damping enhances the nonreciprocal character of the effective damping.
The asymmetry increases (decreases) for the $\omega_-^u$ ($\omega_-^l$) mode. It is maximized at $k_z a = 3\pi/4$, where the odd-in-$\bm{k}$ contributions originating from antisymmetric damping are largest. At this wave vector, the phase difference between neighboring spins maximizes the antisymmetric contribution to the magnetization relaxation dynamics, resulting in the strongest asymmetry in the effective magnon damping.
Furthermore, the asymmetry is identical for both acoustic and optical magnon branches. Notably, the nonlocal Gilbert damping ($\alpha^{AB} \neq 0$) does not contribute to the asymmetry, consistent with the earlier discussion. We further observe that the asymmetry is linear in the long-wavelength limit ($\bm{k}\rightarrow 0$), consistent with the experimental observations~\cite{Kim2024PRL}.

We note that by choosing a nonzero DMI, the system also acquires an effective damping $\alpha_{\rm eff}$ asymmetric with respect to the wave vector in the absence of antisymmetric Gilbert damping.
The corresponding results are presented in Appendix~\ref{AppendixD}. This behavior contrasts with the single-sublattice case, where the effective damping remains insensitive to DMI but depends {\it only} on antisymmetric damping.
Since both antisymmetric Gilbert damping and DMI introduce asymmetry in the effective damping, we compare their respective contributions in Appendix~\ref{AppendixD} and demonstrate that antisymmetric damping induces a much stronger asymmetry. 

Our results provide a practical strategy for experimentally distinguishing the effects of the DMI from those of  antisymmetric damping. In Brillouin-light scattering measurements, the DMI manifests itself through the wave vector dependence of the frequency nonreciprocity, whereas antisymmetric damping is reflected in the asymmetry of the magnon linewidth. It has been reported  recently that the DMI weakens with increasing magnetic-layer thickness, whereas the antisymmetric damping exhibits the opposite trend and becomes significantly enhanced~\cite{Kim2024PRL}. Such contrasting thickness dependencies therefore provide a direct experimental means of disentangling the contributions of DMI and antisymmetric damping. Finally, we mention that no theoretical values exist for the estimation of antisymmetric Gilbert damping, even though the nonlocal Gilbert damping has been computed using first-principles calculations \cite{Nagyfalusi2024,Thonig2018}.

\section{Conclusion}
\label{Sec4}
In this work, we have developed an effective framework to
describe $\bm{k}$-dependent and asymmetric damping in ferromagnetic systems, incorporating both nonlocal Gilbert damping and antisymmetric damping contributions. We first considered a minimal model of a single-sublattice ferromagnet with nearest-neighbor exchange interactions and computed the effective damping.
We have shown that nonlocal damping gives rise to a quadratic dependence of the effective damping on the wave vector, consistent with recent experimental observations, while preserving symmetry with respect to $\bm{k}$. In contrast, the inclusion of antisymmetric Gilbert damping introduces an odd-in-$\bm{k}$ contribution, leading to asymmetric effective damping and, consequently, nonreciprocal damping for counter-propagating magnons. We further did find that the computed damping remains independent of the Dzyaloshinskii-Moriya interaction.

Extending our analysis to two-sublattice ferromagnets on a rectangular lattice structure, we have shown that cross-sublattice antisymmetric damping plays a crucial role in generating asymmetry in the effective damping, whereas nonlocal Gilbert damping does not contribute to such asymmetry.  
Quantifying the degree of magnon damping asymmetry through a dimensionless parameter $\Gamma( \bm{k}) \propto [\alpha_{\rm eff}(\bm{k}) - \alpha_{\rm eff}(-\bm{k})]$ within the Brillouin zone, we have found that this parameter depends strongly on the wave vector $\bm{k}$, reaching a maximum at specific points in reciprocal space. Furthermore, the asymmetry $\Gamma (\bm{k})$ is identical for both acoustic and optical magnon branches.

We also demonstrated that, although DMI can induce an asymmetry in the effective damping of two-sublattice systems, for the parameter regime considered here, its contribution is substantially weaker than that arising from antisymmetric Gilbert damping, with the latter producing an effective asymmetry approximately 25 times larger. This highlights the fundamentally different roles of DMI and antisymmetric damping, with the former primarily modifying the dispersion and the latter directly influencing the dissipative dynamics.
To conclude, our results provide a unified understanding of $\bm{k}$-dependent and nonreciprocal damping in magnetic systems. In particular, our findings are consistent with experimental observations of asymmetric magnon damping in heterostructured magnetic systems, where the DMI becomes very small.
This behavior can be explained through antisymmetric Gilbert damping that can provide an independent microscopic mechanism for nonreciprocal 
magnon dissipation.

\section*{Acknowledgments}
We thank Akashdeep Kamra for fruitful discussions. 
R.M.\ and A.M.\ acknowledge the Indo-German Science Technology Centre (IGSTC) for the exchange visit via the PECFAR award No.\ IGSTC/PECFAR/Call 2024/IGSTC-02232/RM-AM/70/2024-25/74. R.M.\ further acknowledges SERB-SRG via Project No.\ SRG/2023/000612 and the faculty research scheme at IIT (ISM) Dhanbad, India, under Project No.\ FRS(196)/2023-2024/PHYSICS. P.M.O.\ acknowledges financial support from 
the Knut and Alice Wallenberg Foundation (Grants No.\ 2022.0079 and 2023.0336) and from
the European Innovation Council under Pathfinder OPEN Grant Agreement No.\ 101129641 ``OBELIX''. A.M.\ acknowledges funding from the Deutsche Forschungsgemeinschaft (DFG, German Research Foundation) through TRR 173-268565370 (Project B13), and Project No.\ 504261060 (Emmy Noether Programme), and from the Dynamics and Topology Center (TopDyn) funded by the State of Rhineland-Palatinate. 
\label{Sec5}

\section*{DATA AVAILABILITY}
There are no publicly available research data or software
supporting this manuscript. Requests for further information
or data should be sent to the authors.

\appendix
\renewcommand{\thefigure}{\thesection\arabic{figure}}
\onecolumngrid
\section{Derivation of a set of linearized LLG equations for a single-sublattice ferromagnet}
\label{AppendixA}
Here, we provide a detailed derivation of Eq.\ (\ref{Eq5}) in the main text.   
The effective magnetic field $\bm{B}_{i}^{\rm eff}$ that enters the LLG equation can be evaluated using the Hamiltonian \eqref{Eq4} as
\begin{align}
    \bm{B}_{i}^{\rm eff} & = - \frac{1}{M} \frac{\partial \mathcal{H}}{\partial \bm{\sigma}_{i}} = - \frac{1}{M} \frac{\partial \mathcal{H}}{\partial {\beta}_{2i}}\hat{\bm{y}} + \frac{1}{M} \frac{\partial \mathcal{H}}{\partial {\beta}_{1i}}\hat{\bm{z}}.
    \label{App_A_Eq1}
\end{align}
 We use this effective magnetic field in the LLG spin dynamics together with scalar, $\alpha_{ij}$, and antisymmetric, $\bm{C}_{ij}$, damping. Within the LLG formalism, the decoupled angle variables take the following form
\begin{align}
\frac{d\beta_{2i}}{dt} &=  \frac{\gamma_i}{M_i}\frac{\partial \mathcal{H}}{\partial {\beta}_{1i}} +\sum_{j}\alpha_{ij}\frac{d \beta_{1j}}{dt}+ \sum_{i \neq j}C_{ij}\frac{d \beta_{2j}}{dt}, \label{App_A_Eq2}\\
\frac{d\beta_{1i}}{dt} &=  -\frac{\gamma_i}{M_i}\frac{\partial \mathcal{H}}{\partial {\beta}_{2i}} -\sum_{j}\alpha_{ij}\frac{d \beta_{2j}}{dt}+ \sum_{i \neq j}C_{ij}\frac{d \beta_{1j}}{dt}. \label{App_A_Eq3}
\end{align}

Next, the following derivatives are evaluated using the Hamiltonian of Eq.\ \eqref{Eq4} as follows
\begin{align}
\frac{\partial\mathcal{H}}{\partial\beta_{2i}} &= \sum_{j}\left( J^{yy}_{ij}\beta_{2j}  - J^{xx}_{ij}\beta_{2i}\right) - \sum_{j}D_{ij}^x \beta_{1j}  - 2K^x\beta_{2i} + B_{i}^x \beta_{2i} M_{i} - B_{i}^yM_{i}
\label{App_A_Eq5},\\
\frac{\partial\mathcal{H}}{\partial\beta_{1i}} &= \sum_{j}\left( J^{zz}_{ij}\beta_{1j}  - J^{xx}_{ij}\beta_{1i}\right) + \sum_{j}D_{ij}^x \beta_{2j} - 2K^x\beta_{1i} + B_{i}^x \beta_{1i} M_{i} + B_{i}^zM_{i}
\label{App_A_Eq6}.
\end{align}
Further, we use the following Fourier transformations and obtain $\bm{k}$-dependent angle variables and magnetic fields
\begin{align}
\tilde{\beta}_{1(2)}(\bm{k})  = \frac{1}{\sqrt{N}} \sum_{\bm{R}_i} e^{-{\rm i} \bm{k}\cdot \bm{R}_i} \beta_{1j(2j)}\, , \quad
\tilde{B}^{y(z)}(\bm{k})  = \frac{1}{\sqrt{N}} \sum_{\bm{R}_i} e^{-{\rm i} \bm{k}\cdot \bm{R}_i} B^{y(z)}_i\,.
\end{align}
We assume that, for $i=j$, $\alpha_{ii}=\alpha$, corresponding to the conventional local Gilbert damping. For $i\neq j$, we construct the nonlocal damping parameter $\alpha_{ij}$ by considering only nearest-neighbor contributions. Furthermore, since $\bm{\mathcal{C}}_{ij}$ is a vector quantity defined between distinct lattice sites, it vanishes for $i=j$. In $\bm{k}$-space, the Eqs.\ \eqref{App_A_Eq2} and \eqref{App_A_Eq3} can be written as
\begin{align}
\frac{d\tilde{\beta_2}(\bm{k})}{dt} &= \frac{\gamma}{M}\Bigg[\tilde{J}^{zz}(\bm{k})\tilde{\beta_1}(\bm{k}) - {\Omega}\tilde{\beta_1}(\bm{k}) +\tilde{D}(\bm{k})\tilde{\beta}_2(\bm{k}) + \tilde{B}^z(\bm{k})M \Bigg] + \alpha \frac{d \tilde{\beta}_1(\bm{k})}{dt} + \tilde{\alpha}(\bm{k})\frac{d\tilde{\beta}_1(\bm{k})}{dt} + \tilde{C}(\bm{k})\frac{d\tilde{\beta}_2(\bm{k})}{dt}\label{App_A_Eq7},\\
\frac{d\tilde{\beta_1}(\bm{k})}{dt} &= \frac{\gamma}{M}\Bigg[-\tilde{J}^{yy}(\bm{k})\tilde{\beta_2}(\bm{k}) + {\Omega}\tilde{\beta_2}(\bm{k}) +\tilde{D}(\bm{k})\tilde{\beta}_1(\bm{k}) +\tilde{B}^y(\bm{k})M \Bigg]- \alpha \frac{d \tilde{\beta}_2(\bm{k})}{dt} - \tilde{\alpha}(\bm{k})\frac{d\tilde{\beta}_2(\bm{k})}{dt}  + \tilde{C}(\bm{k})\frac{d\tilde{\beta}_1(\bm{k})}{dt}.
\label{App_A_Eq8}
\end{align}
In these equations, tilde signs (\,\,$\tilde{\empty}$\,\,) refer to the Fourier transformed counterparts. The above occurring quantities are computed as
\begin{align}
     \sum_mz_mJ^{xx}_m + 2K^x  -B^xM& = {\Omega} \,,\\
    \sum_{j} J^{zz} e^{{\rm i}\bm{k}\cdot\bm{R_j}}e^{{\rm -i}\bm{k}\cdot\bm{R_i}} = \sum_{\bm{\delta}} J^{zz} e^{{\rm i}\bm{k}\cdot\bm{\delta}} & = \tilde{J}^{zz}(\bm{k}),\\
    \sum_{j} D_{ij} e^{{\rm i}\bm{k}\cdot\bm{R_j}}e^{{\rm -i}\bm{k}\cdot\bm{R_i}} = \sum_{\bm{\delta}} D_{ij} e^{{\rm i}\bm{k}\cdot\bm{\delta}} & = \tilde{D}(\bm{k}),\\
    \sum_{j} \alpha_{ij} e^{{\rm i}\bm{k}\cdot\bm{R_j}}e^{{\rm -i}\bm{k}\cdot\bm{R_i}} = \sum_{\bm{\delta}} \alpha_{ij} e^{{\rm i}\bm{k}\cdot\bm{\delta}} & = \tilde{\alpha}(\bm{k}), \label{App_A_Eq9}\\
    \sum_{j} C_{ij} e^{{\rm i}\bm{k}\cdot\bm{R_j}}e^{{\rm -i}\bm{k}\cdot\bm{R_i}} = \sum_{\bm{\delta}} C_{ij} e^{{\rm i}\bm{k}\cdot\bm{\delta}} & = \tilde{C}(\bm{k}).
    \label{App_A_Eq10}
\end{align}
Here $z_m$ is the number of $m$-th nearest neighbors and $\bm{\delta}$ represents a vector connecting $i$-th and $j$-th lattice sites, being nearest neighbors. Assuming ${J}^{yy} = {J}^{zz}$, the Eqs.~\eqref{App_A_Eq7} and ~\eqref{App_A_Eq8} combine in the circular basis [$\tilde\beta_{\pm}(\bm{k}) = \tilde\beta_2(\bm{k}) \pm {\rm i}\tilde\beta_1(\bm{k})$] as
\begin{align}
    \frac{d\tilde{\beta}_{\pm}(\bm{k})}{dt} &= \frac{\gamma}{M}\Big[\mp{\rm i}\tilde{J}^{zz}(\bm{k})\tilde{\beta}_{\pm}(\bm{k}) \pm{\rm i}\tilde{\Omega}\tilde{\beta}_{\pm}(\bm{k})+\tilde{D}(\bm{k})\tilde{\beta}_{\pm}(\bm{k}) + \tilde{B}_{\pm}(\bm{k})M \Big] \mp{\rm i} \alpha \frac{d \tilde{\beta}_{\pm}(\bm{k})}{dt} \mp{\rm i} \tilde{\alpha}(\bm{k}) \frac{d \tilde{\beta}_{\pm}(\bm{k})}{dt} + \tilde{C}(\bm{k}) \frac{d \tilde{\beta}_{\pm}(\bm{k})}{dt}.
    \label{App_A_Eq11}
\end{align}
For small deviation from the ground state, we can assume $\tilde{\beta}_{\pm} \propto e^{\pm {\rm i}\omega t}$. Then, the above equation simplifies to:
\begin{align}
\left[{ \pm {\rm i}\omega\mathcal{A}({\bm k})} + {\mathcal{J}}({\bm k})\right]\tilde\beta_{\pm}(\bm{k}) = \gamma B_{\pm}(\bm{k})\,,
\label{App_A_Eq13}
\end{align}
where the corresponding $\bm{k}$-dependent coefficients are
\begin{align}
    \mathcal{A}(\bm{k}) &= 1 \pm {\rm i}\alpha \pm {\rm i}\tilde{\alpha}(\bm{k}) - \tilde{C}(\bm{k}), \\
    \mathcal{J}(\bm{k}) &= \frac{\gamma}{M} \Big[\pm {\rm i}\tilde{J}^{zz}(\bm{k}) \mp {\rm i} \tilde{\Omega} -\tilde{D}(\bm{k}) \Big].
\end{align}

\section{Derivation of a set of linearized LLG equations for a two-sublattice ferromagnet}
\label{AppendixC}
Here, we provide the detailed derivation of Eq.~\eqref{Eq15}. The two sublattices are denoted as $p\in \{A,B\}$. Similar to the one-sublattice case, the decoupling of the LLG equation in Eq.~\eqref{Eq13} in $y$ and $z$ components, we get the following set of equations:
\begin{align}
\frac{d\beta_{2i_p}^p}{dt} &=  \frac{\gamma^p}{M^p}\frac{\partial \mathcal{H}}{\partial {\beta}_{1i_p}^p} +\sum_q\sum_{j_q}\alpha_{i_pj_q}^{pq}\frac{d \beta_{1j_q}^q}{dt} + \sum_q\sum_{i_p \neq j_q}C^{pq}_{i_pj_q}\frac{d \beta_{2j_q}^q}{dt}\label{App_B_Eq1} ,\\
\frac{d\beta_{1i_p}^p}{dt} &=  -\frac{\gamma^p}{M^p}\frac{\partial \mathcal{H}}{\partial {\beta}_{2i_p}^p} -\sum_q\sum_{j_q}\alpha_{i_pj_q}^{pq}\frac{d \beta_{2j_q}^q}{dt} + \sum_q\sum_{i_p \neq j_q}C^{pq}_{i_pj_q}\frac{d \beta_{1j_q}^q}{dt} .
\label{App_B_Eq2}
\end{align}
The Hamiltonian terms in Eq.~\eqref{Eq14} can be written in terms of angle variables as
\begin{align}
    \mathcal{H}  &= \frac{1}{2}\sum_{p,q}\sum_{\langle i_p,j_q\rangle} \left[J^{xx,pq}_{i_pj_q}\left\{\left(1 - \dfrac{({\beta}_{1i_p}^{p})^2}{2} - \dfrac{({\beta}_{2i_p}^{p})^2}{2}\right)  \left(1 - \dfrac{({\beta}_{1j_q}^q)^2}{2} - \dfrac{({\beta}_{2j_q}^{q})^2}{2}\right)\right\} + J^{yy,pq}_{i_p,j_q} \beta_{2i_p}^p\beta_{2j_q}^q+J^{zz,pq}_{i_pj_q}\beta_{1i_p}^p\beta_{1j_q}^q \right] \nonumber\\ & \quad + \sum_{p,q}\sum_{\langle i_p,j_q \rangle}\varepsilon^{pq}D^{x}_{i_pj_q}\beta_{1i_p}^p\beta_{2j_q}^q + K^{x,p} \sum_{i_p}\left(1 - \dfrac{({\beta}_{1i_p}^{p})^2}{2} - \dfrac{({\beta}_{2i_p}^{p})^2}{2}\right)  - \sum_{i_p}\Bigg[ B_{i_p}^x\left(1 - \dfrac{({\beta}_{1i_p}^{p})^2}{2} - \dfrac{({\beta}_{2i_p}^{p})^2}{2}\right)M^{p}_{i_p} \nonumber\\& \quad + B_{i_p}^y\beta_{2i_p}^{p}M^{p}_{i_p}  - B_{i_p}^z\beta_{1i_p}^{p}M^{p}_{i_p} \Bigg],\label{App_B_Eq3}\end{align}
where
\begin{align}
\varepsilon^{pq} =
\begin{pmatrix}
    0 & 1\\
    -1 & 0
\end{pmatrix}.
\end{align}
 Now, the derivatives in Eqs.\ \eqref{App_B_Eq1} and \eqref{App_B_Eq2} are evaluated as follows
\begin{align}
\frac{\partial\mathcal{H}}{\partial\beta_{2i_p}^{p}} &= \sum_q\sum_{\langle j_q\rangle}\left( J^{yy,pq}_{i_pj_q}\beta_{2j_q}^q  - J^{xx,pq}_{i_pj_q}\beta_{2i_p}^{p}\right) - \sum_q\sum_{\langle j_q \rangle}D^{x,pq}_{i_pj_q}\beta_{1j_q}^q - 2K^{x,p}\beta_{2i_p}^p + B_{i_p}^x \beta_{2i_p}^p M^p_{i_p} - B_{i_p}^yM^{p}_{i_p}
\label{App_B_Eq4},\\
\frac{\partial\mathcal{H}}{\partial\beta_{1i_p}^{p}} &= \sum_q\sum_{\langle j_q\rangle}\left( J^{zz,pq}_{i_pj_q}\beta_{1j_q}^q  - J^{xx,pq}_{i_pj_q}\beta_{1i_q}^{p}\right) + \sum_q\sum_{\langle j_q \rangle}D^{x,pq}_{i_pj_q}\beta_{2j_q}^q - 2K^{x,p}\beta_{1i_p}^p + B_{i_p}^x \beta_{1i_p}^p M^p_{i_p} + B_{i_p}^zM^{p}.
\label{App_B_Eq5}
\end{align}

In  $\bm{k}$-space, Eqs.\ \eqref{App_B_Eq1} and \eqref{App_B_Eq2} can be written as
\begin{align}
 \frac{d \tilde{\beta}_2^p(\bm{k})}{dt} & = \frac{\gamma^p}{M^p}\Biggl[\sum_q\tilde{J}^{zz,pq}(\bm{k})\tilde{\beta_1}^q(\bm{k}) - {\Omega}^p\tilde{\beta_1}^p(\bm{k}) +  \sum_{q}\tilde{D}^{pq}(\bm{k})\tilde{\beta_2}^q(\bm{k}) + B^zM^p \Biggl] + \sum_q\tilde{\alpha}^{pq}\frac{d\tilde{\beta_1}^q(\bm{k})}{dt} +\sum_q\tilde{C}^{pq}\frac{d\tilde{\beta_2}^q(\bm{k})}{dt}, \label{App_B_Eq6}\\
 \frac{d \tilde{\beta}_1^p(\bm{k})}{dt} & = \frac{\gamma^p}{M^p}\Biggl[-\sum_q\tilde{J}^{yy,pq}(\bm{k})\tilde{\beta_2}^q(\bm{k}) + {\Omega}^p\tilde{\beta_2}^p(\bm{k}) +  \sum_{q}\tilde{D}^{pq}(\bm{k})\sigma_1^{pq}\tilde{\beta_1}^q(\bm{k}) + B^yM^p \Biggl] - \sum_q\tilde{\alpha}^{pq}\frac{d\tilde{\beta_2}^q(\bm{k})}{dt} +\sum_q\tilde{C}^{pq}\frac{d\tilde{\beta_1}^q(\bm{k})}{dt},
 \label{App_B_Eq7}
\end{align}
where the Fourier transformed quantities are
\begin{figure*}[tbh!]
    \centering
\includegraphics[width=0.7\linewidth]{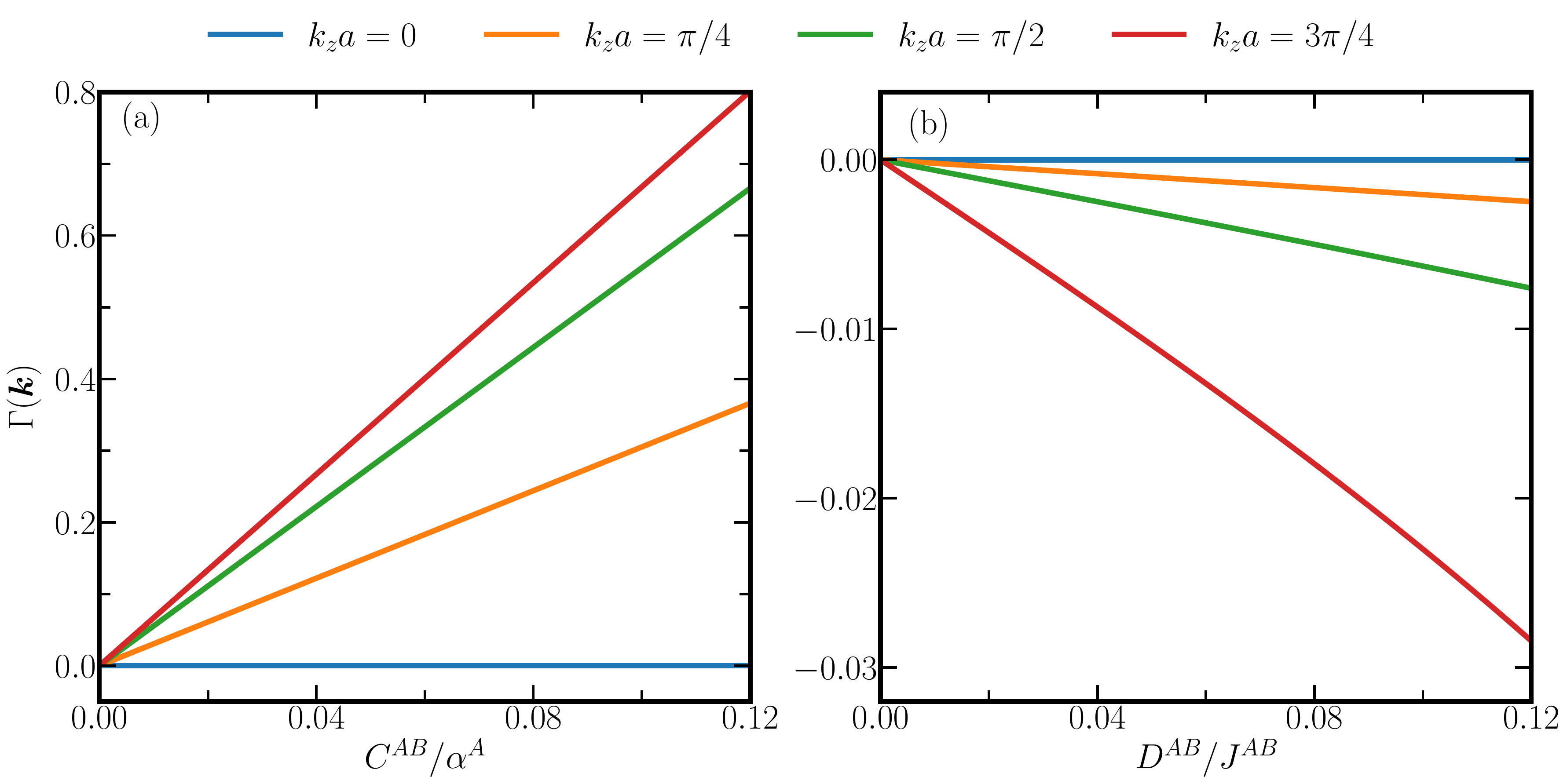}
    \caption{The computed asymmetry parameter $\Gamma (\bm{k})$ as a function of (a) antisymmetric  Gilbert damping, and (b) DMI at specific $\bm{k}$-points within the Brillouin zone while keeping $k_x = 0$. The effect of the antisymmetric damping parameter is much larger than that of the DMI.
    }
    \label{fig:Appendix}
\end{figure*}
\begin{align}
       \sum_{\langle j_q \rangle} J^{yy/zz,pq} e^{{\rm i}\bm{k}\cdot\bm{R}_{j_q}^q}e^{{\rm -i}\bm{k}\cdot\bm{R}_{i_p}^p} &= \sum_{\bm{\delta}^{pq}} J^{yy/zz,pq} e^{{\rm i}\bm{k}\cdot\bm{\delta}^{pq}} = \tilde{J}^{yy/zz, pq}(\bm{k}),\\
    \sum_{\langle j_q \rangle} D^{x,pq} e^{{\rm i}\bm{k}\cdot\bm{R}_{j_q}^q}e^{{\rm -i}\bm{k}\cdot\bm{R}_{i_p}^p} &= \sum_{\bm{\delta}^{pq}} D^{x,pq} e^{{\rm i}\bm{k}\cdot\bm{\delta}^{pq}} = \tilde{D}^{pq}(\bm{k}),\\
    \sum_{\langle j_q \rangle} \alpha_{i_pj_q}^{pq} e^{{\rm i}\bm{k}\cdot\bm{R}_{j_q}^q}e^{{\rm -i}\bm{k}\cdot\bm{R}_{i_p}^p} &= \sum_{\bm{\delta}^{pq}} \alpha_{i_pj_q}^{pq} e^{{\rm i}\bm{k}\cdot\bm{\bm{\delta}}^{pq}} =\tilde{\alpha}^{pq}(\bm{k}),\\
    \sum_{\langle j_q \rangle} {C}_{i_pj_q}^{pq} e^{{\rm i}\bm{k}\cdot\bm{R}_{j_q}^q}e^{{\rm -i}\bm{k}\cdot\bm{R}_{i_p}^p} &= \sum_{\bm{\delta}^{pq}} C_{i_pj_q}^{pq} e^{{\rm i}\bm{k}\cdot\bm{\bm{\delta}}^{pq}} = \tilde{C}^{pq}(\bm{k})\,,
\end{align}
with ${\Omega}^p = \sum_{q,\langle j_q\rangle }z_{j_q}{J}^{xx,pq}_{i_p,j_q} + 2K_x^p - B_{x}M^p $,  $\bm{\delta}^{pq}$ is the lattice vector connecting sublattice $p$ and $q$, and $z_{j_q}$ is the number of $j_q$-th nearest neighbors. In the circularly polarized basis, equations \eqref{App_B_Eq6} and \eqref{App_B_Eq7} can  be combined as:
\begin{align}
\frac{d\tilde{\beta}_\pm ^p(\bm{k})}{dt} &= \frac{\gamma^p}{M^p}\Biggl[\mp \sum_q {\rm i}\tilde{J}^{pq}(\bm{k}) \tilde{\beta}^q_{\pm} \pm {\rm i} {\Omega}^p \tilde{\beta}^p_{\pm} +\sum_q\tilde{D}^{pq}(k)\tilde{\beta}^q_{\pm}
+\tilde{B}_{\pm}M^p \Biggl] \mp \sum_q {\rm i} \tilde\alpha^{pq}(\bm{k}) \frac{d\tilde{\beta}_{\pm}^q(\bm{k})}{dt} + \sum_q\tilde{C}^{pq}(\bm{k})\frac{d\tilde{\beta}_{\pm}^q(\bm{k})}{dt}.
    \label{App_B_Eq14}
\end{align}
Here, we have assumed $\tilde{J}^{yy,pq} = \tilde{J}^{zz,pq} = \tilde{J}^{pq}$. We restrict all interactions in our model up to the first nearest neighbors. For small deviations from the ground state and considering two sublattices $A$ and $B$, Eq.\ \eqref{App_B_Eq14} simplifies to
\begin{align}
\left[{\pm \rm{i}\omega\mathcal{A}_2({\bm k})} + {\mathcal{J}}_2({\bm k})\right]\tilde\beta_{\pm}(\bm{k}) = \mathcal{B}_{\pm}(\bm{k}).
\label{App_B_Eq19}
\end{align}
where,
\begin{align}
    \mathcal{A}_2(\bm{k}) = 
    \begin{pmatrix}
1 \pm {\rm i} \alpha^{A} &  \pm {\rm i}\tilde{\alpha}^{AB}(\bm{k}) - \tilde{C}^{AB}(\bm{k})\\\\ \pm {\rm i} \tilde{\alpha}^{BA} (\bm{k})  - \tilde{C}^{BA}(\bm{k}) &
1 \pm {\rm i} \alpha^{B}
\end{pmatrix},
\end{align}
\begin{align}
    \mathcal{J}_2(\bm{k}) = 
\begin{pmatrix}
\mp {\rm i} \dfrac{\gamma^A}{M^A} \tilde{\Omega}_A  & {\rm i}\dfrac{\gamma^A}{M^A} \bigg( \pm \tilde{J}^{AB}(\bm{k}) + {\rm i}\tilde{D}^{AB}(\bm{k}) \bigg)\\\\{\rm i}\dfrac{\gamma^B}{M^B} \bigg( \pm \tilde{J}^{BA}(\bm{k}) + {\rm i} \tilde{D}^{BA}(\bm{k}) \bigg) &
 \mp {\rm i} \dfrac{\gamma^B}{M^B} \tilde{\Omega}_B
\end{pmatrix} ,
\end{align}
\begin{align}
\tilde\beta_{\pm} = 
\begin{pmatrix}
\\
\tilde{\beta}_{\pm}^A(\bm{k}) \\\\
\tilde{\beta}_{\pm}^B(\bm{k})\\\\
\end{pmatrix}, \quad \quad
\mathcal{B}_\pm = 
\begin{pmatrix}\\
\gamma^A\tilde{B}_{\pm}\\\\
\gamma^B\tilde{B}_{\pm}\\\\
\end{pmatrix}.
\end{align}
Note that this formalism is general for a two-sublattice ferromagnet. This derived result is Eq.~\eqref{Eq15} in the main text.

\section{Comparison of  the asymmetry in the effective damping}
\label{AppendixD}  
For a two-sublattice ferromagnet, we compute the asymmetry $\Gamma(\bm{k})$  according to Eq.~\eqref{Eq15n}. We distinguish the effect of antisymmetric Gilbert damping and the DMI and calculated $\Gamma(\bm{k})$ for 
the $\omega_-^u$ mode only; the results are presented in Fig.\ \ref{fig:Appendix}. The presented calculations signify that the antisymmetric damping is the strongest mechanism for producing effective asymmetric damping, as compared to the DMI.

\twocolumngrid
\bibliography{Reference}

\end{document}